\documentclass[aps,pra,nofootinbib,amsmath,amssymb,superscriptaddress,twocolumn]{revtex4-2}

\usepackage[utf8]{inputenc}
\usepackage[T1]{fontenc}

\usepackage{graphicx}
\usepackage{xcolor}

\usepackage[english]{babel}
\usepackage{verbatim}
\usepackage{soul}

\usepackage[separate-uncertainty=true]{siunitx}
\usepackage[version=3]{mhchem}

\usepackage{hyperref}
\usepackage[full]{textcomp}
\usepackage{mathtools}

\usepackage{etoolbox}
\usepackage{tikz}
\usepackage{stix}

\newrobustcmd*{\mycircle}[1]{\tikz{\filldraw[draw=#1,fill=#1] (0,0) circle [radius=0.1cm];}}

\newrobustcmd*{\mytriangle}[1]{\tikz{\filldraw[draw=#1,fill=#1] (0,0) --
(0.2cm,0) -- (0.1cm,0.2cm);}}

\newrobustcmd*{\mytriangleUSD}[1]{\tikz{\filldraw[draw=#1,fill=#1] (0,0.2cm) --
(0.2cm,0.2cm) -- (0.1cm,0cm);}}

\newrobustcmd*{\mysquare}[1]{\tikz{\filldraw[draw=#1,fill=#1] (0,0) --
(0,0.2cm) -- (0.2cm,0.2cm) -- (0.2cm,0cm);}}
\newcommand{\COMM}[1]{\iffalse {#1} \fi}

\newcommand{\highlightingOn}{0}

\if\highlightingOn1
\definecolor{hL1}{rgb}{1,0,0}
\definecolor{hL2}{rgb}{0.7,0,0}
\definecolor{hL3}{rgb}{0.45,0,0.8}
\else
\definecolor{hL1}{rgb}{0,0,0}
\definecolor{hL2}{rgb}{0,0,0}
\definecolor{hL3}{rgb}{0,0,0}
\fi

\newcommand{\firstrevision}[1]{\textcolor{hL1}{#1}}
\newcommand{\secondrevision}[1]{\textcolor{hL2}{#1}}
\newcommand{\thirdrevision}[1]{\textcolor{hL3}{#1}}

\usepackage{tikz,xcolor}
\definecolor{lime}{HTML}{A6CE39}
\DeclareRobustCommand{\orcidicon}{%
	\begin{tikzpicture}
	\draw[lime, fill=lime] (0,0) 
	circle [radius=0.16] 
	node[white] {{\fontfamily{qag}\selectfont \tiny ID}};
	\draw[white, fill=white] (-0.0625,0.095) 
	circle [radius=0.007];
	\end{tikzpicture}
	\hspace{-2mm}
}
\foreach \x in {A, ..., Z}{%
	\expandafter\xdef\csname orcid\x\endcsname{\noexpand\href{https://orcid.org/\csname orcidauthor\x\endcsname}{\noexpand\orcidicon}}
}

\begin{document}

\author{Lukas Tenbrake\orcidE{}}
 \thanks{These authors contributed equally to this work.}
 \affiliation{Institute of Applied Physics, University of Bonn, Germany}

\author{Alexander Faßbender}
 \thanks{These authors contributed equally to this work.}
 \affiliation{Institute of Physics, University of Bonn, Germany}

\author{Sebastian Hofferberth\orcidB{}}
 \affiliation{Institute of Applied Physics, University of Bonn, Germany}

\author{Stefan Linden\orcidC{}}
 \affiliation{Institute of Physics, University of Bonn, Germany}

\author{Hannes Pfeifer\orcidD{}}
 \altaffiliation{hannes.pfeifer@iap.uni-bonn.de,\newline Current address: Department of Microtechnology and Nanoscience, Chalmers University of Technology, Gothenburg, Sweden}
 \affiliation{Institute of Applied Physics, University of Bonn, Germany}

\title{Direct laser-written optomechanical membranes in fiber Fabry-Perot cavities}

\date{\today}

\begin{abstract}
\COMM{Recent m}Integrated micro- and nanophotonic optomechanical experiments \COMM{achieving record optomechanical coupling strengths }enable\COMM{d} the manipulation of mechanical resonators on the single phonon level. Interfacing these structures requires elaborate techniques limited in tunability, flexibility, and scaling towards multi-mode systems. Here, we demonstrate a cavity optomechanical experiment using 3D-laser-written polymer membranes inside fiber Fabry-Perot cavities. Vacuum coupling rates of $g_0/2\pi \approx \SI{30}{\kilo\hertz}$ to the fundamental megahertz mechanical mode are reached. We observe optomechanical spring tuning of the mechanical resonator frequency by tens of kilohertz exceeding its linewidth at cryogenic temperatures.\COMM{ The extreme flexibility of the laser writing process allows for a direct integration of the membrane into the microscopic cavity.} The direct fiber coupling, its scaling capabilities to coupled resonator systems, and the potential implementation of dissipation dilution structures and integration of electrodes make it a promising platform for fiber-tip integrated accelerometers, optomechanically tunable multi-mode mechanical systems, and directly fiber-coupled systems for microwave to optics conversion. 
\end{abstract}

\maketitle

\section{Introduction}
Cavity optomechanical experiments have been implemented on a multitude of different platforms \cite{aspelmeyer2014cavity} ranging from the canonical movable end mirror of a Fabry-Perot cavity \cite{groblacher2009observation}, over membranes in cavities \cite{thompson2008strong, jayich2008dispersive}, toroidal\cite{schliesser2008resolved} and optomechanical crystal resonators \cite{eichenfield2009optomechanical}, down to the vibrational modes of molecules in a plasmonic picocavity \cite{benz2016single}. This platform diversification and constant improvement advanced the field during the past years leading to key achievements like ground-state cooling of mechanical resonators \cite{chan2011laser, teufel2011sideband}, optomechanical state teleportation experiments \cite{fiaschi2021optomechanical}, efficient microwave to optical conversion \cite{andrews2014bidirectional, han2021microwave},\COMM{ and preparation of non-classical mechanical states \cite{zivari2022non}, squeezing of both light and motion \cite{safavi2013squeezed, purdy2013strong, wollman2015quantum, pirkkalainen2015squeezing}} or sensing of the mechanical resonator below the standard quantum limit \cite{teufel2009nanomechanical, anetsberger2010measuring, mason2019continuous}. Among the current challenges in the field are the realization and addressing of multi-mode optomechanical systems and the integration of optomechanical elements for different sensing applications.

Optomechanical devices with high optical field concentration and correspondingly large optomechanical coupling are usually realized with on-chip platforms \cite{eichenfield2009optomechanical, schliesser2008resolved, zhang2012synchronization}. A different approach is taken by miniaturized Fabry-Perot cavities with concave mirror structures fabricated on optical fiber tips \cite{hunger2010fiber, pfeifer2022achievements}. These fiber Fabry-Perot cavities (FFPCs) have been established as a platform for light-matter interaction during the past years, including \firstrevision{experiments on atoms inside FFPCs for photonic qubits or quantum networks \cite{brekenfeld2020quantum, niemietz2021nondestructive}, and} realizations of optomechanical experiments \cite{flowers2012fiber, shkarin2014optically, kashkanova2017superfluid, kashkanova2017optomechanics, shkarin2019quantum, fogliano2021mapping, rochau2021dynamical}. They feature a direct fiber-coupled optical access, small cavity lengths, high optical finesse, and an open resonator volume. This allows to introduce both conventional membrane-type resonators as well as more unconventional resonators like standing waves in liquid Helium. 

With its first demonstration in the 1990s \cite{maruo1997three}, 3D direct laser writing (DLW) enabled the fabrication of free-formed three- dimensional polymeric structures with sub-micrometer resolution. It led to the miniaturization of on-chip optical components \cite{gissibl2016two}, waveguides \cite{jorg2020artificial}, and mechanical structures \cite{williams2020dynamic} including 3D acoustic metamaterials \cite{frenzel2017three}, and has been used for mask applications \cite{pfeifer2022achievements}. Apart from large planar substrates, writing on fiber-ends has been tackled for applications in endoscopy \cite{li2020ultrathin} or sensing \cite{thompson2018micro, zou2021fiber}. 

In this article, we demonstrate the integration of mechanical polymer membrane resonators into highly miniaturized FFPCs. We realize a miniaturized, fiber-coupled membrane-in-the-middle (MIM) experiment \cite{thompson2008strong, jayich2008dispersive} with superior scaling capabilities due to the 3D DLW fabrication process both integrated into highly stable monolithic FFPCs \cite{saavedra2021tunable} and on flat distributed Bragg reflector (DBR) substrates. We analyze the achievable optomechanical coupling strength, and demonstrate a dispersive optomechanical spring effect tuning of the mechanical resonace in the presence of a thermal optical nonlinearity exceeding the mechanical linewidth at cryogenic temperatures. \firstrevision{To characterize the basic internal material properties no elaborate dissipation dilution structures were used in this first proof-of-principle study.}

Our results pave the way for \firstrevision{using highly flexible DLW structure fabrication for optomechanical resonators. DLW enables new realizations of} high-sensitivity fiber-cavity integrated accelerometers, and mechanical multi-mode structures that are interfaced using FFPCs, including multiple membranes inside a miniaturized Fabry-Perot cavity \cite{xuereb2012strong} \firstrevision{(see methods). Furthermore, extended mechanical metamaterials of membrane resonators on DBR substrates for controlling vibrations in a thin film can be realized and combination with other light-matter interfaces as 2D materials and quantum emitters can be envisaged.}

\section{Results}

\subsection*{Polymer membrane in a fiber cavity}

\begin{figure*}[ht]
\centering
\includegraphics[width=\textwidth]{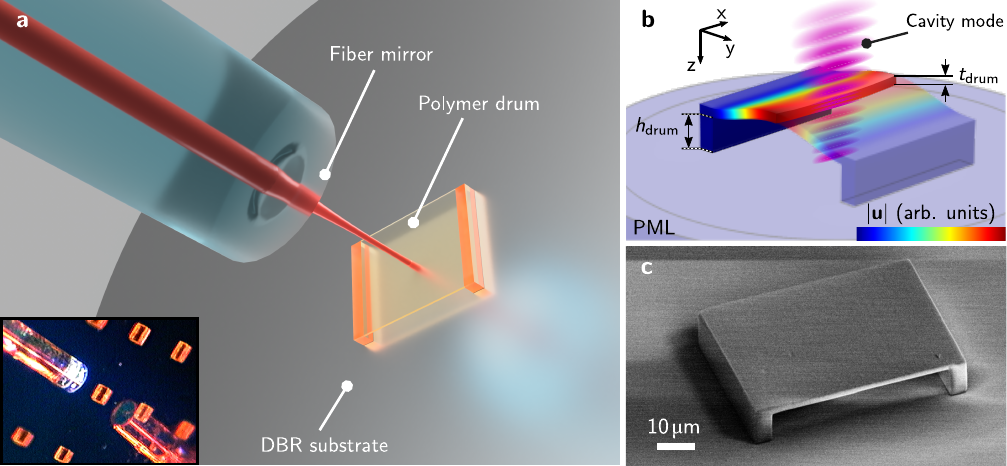}
\caption{Overview of the system. In \textbf{a} a schematic overview of the experimental structure is shown. The polymer drum resonator is fabricated on a highly reflective DBR substrate. The fiber mirror is positioned above the polymer drum and realizes a fiber Fabry-Perot cavity with the DBR beneath enclosing the drum membrane. The inset shows a microscope picture of an approaching fiber mirror (diameter: $\SI{125}{\micro\meter}$) to a polymer drum array. In \textbf{b} the magnitude of the displacement field $\mathbf{u}$ of the fundamental drum mode is shown as retrieved from finite element simulations. The height $h_\text{drum}$ of the polymer drum supports and the drum membrane thickness $t_\text{drum}$ determine the positions of the optical cavity mode intensity maxima and minima with respect to the membrane interfaces. In \textbf{c} an SEM micrograph of a fabricated rectangular polymer drum structure is shown.}
\label{fig:fig01_generalSetup}
\end{figure*}

The mechanical resonator in our experiments is a drum-like polymer membrane of $1-\SI{2}{\micro\meter}$ thickness. It is fabricated using 3D direct laser writing (see methods) on top of a highly reflective DBR mirror ($\SI{10}{ppm}$ transmission\footnote{\label{fn:coatingSpan}High reflectivity range: $\SI{750}{\nano\meter}$ to $\SI{805}{\nano\meter}$}) located either on an extended substrate or on an optical fiber-tip. A typical rectangular membrane used here spans $\sim \SI{45}{\micro\meter}\times \SI{60}{\micro\meter}$ and is placed on support bars that suspend the membrane $5-\SI{10}{\micro\meter}$ above the mirror surface (see Fig.~\ref{fig:fig01_generalSetup}~a and c). The supports reduce the free membrane surface by $\SI{5}{\micro\meter}$ on each side. More elaborate geometries like multi-membrane structures or suspensions like soft-clamping \cite{tsaturyan2017ultracoherent} can be directly realized in the 3D DLW fabrication. Finite-element simulations of the fundamental mechanical mode, see Fig.~\ref{fig:fig01_generalSetup}~b and methods section, yield an expected mechanical resonance frequency of $\SI{2.1}{\mega\hertz}$ and an internal quality factor of $\sim 20$ due to the comparably large elastic loss tangent of the 3D DLW polymer resist material at ambient conditions. As no specific dissipation dilution is implemented in this simple design, the mechanical quality factor reflects the intrinsic $Q$ value \firstrevision{($Q_\text{int}$)} of the material \cite{fedorov2019generalized}.  

The membrane is integrated into an optical FFPC by approaching a fiber mirror -- a fiber-tip with a concave-shaped facet and high-reflection coated surface \cite{hunger2010fiber,pfeifer2022achievements}. The cavity length is adjusted to $\sim\SI{30}{\micro\meter}$, but can be scanned using a piezo-electric element attached to the fiber mirror. The optical fiber leading to the approached fiber mirror serves as both input and output of the cavity. The transmission\footref{fn:coatingSpan} of $\SI{2000}{ppm}$ is chosen to retrieve a single-sided cavity geometry, with an approximate balance of the cavity-to-input-fiber coupling rate and the internal cavity losses. The internal cavity losses are dominated by scattering from the polymer surface that exhibits a roughness\footnote{RMS surface profile variation in AFM measurement} of $< \SI{5}{\nano\meter}$ after post-fabrication polishing using an oxygen plasma ashing process. As the air-polymer interfaces of the membrane have a large transmittance, the field of the optical cavity extends through the membrane and over both open cavity domains.\COMM{The optical cavity mode field extends over the membrane and the open cavity domains on both sides of the membrane due to its low reflectivity.} The optical cavity mode spectrum is characterized by scanning the cavity resonance over a probe laser tone with modulated sidebands (see methods, and \cite{gallego2016high, saavedra2021tunable}). We measure optical linewidths of $\kappa/2\pi$ of few gigahertz corresponding to finesse values of $\mathcal{F}= \SI{1400\pm 300}{}$ for the case of an intensity maximum on one of the polymer-air interfaces and a minimum on the other. As scattering from the membrane surface is the dominant optical loss mechanism, the finesse reaches the empty cavity value for the case of intensity minima on both interfaces. \COMM{We measure optical linewidths of $\kappa/2\pi\sim\SI{}{\giga\hertz}$ corresponding to finesse values of $\approx 1000$. The variations are caused by different optical intensities on the membrane surfaces depending on the individual membrane geometry.}

In order to probe the mechanical mode, we lock the cavity to the probe laser using a feedback loop on a Pound-Drever-Hall (PDH) error signal \cite{drever1983laser, black2001introduction}. The thermal excitation of the membrane is transduced to cavity frequency noise visible in the noise spectrum of the calibrated PDH error signal, which is recorded using an electrical spectrum analyzer. The measured mechanical frequencies and linewidths are in good agreement with the finite element simulation with mechanical resonance frequencies between 1 and $\SI{4}{\mega\hertz}$ depending on the particular membrane geometry. The calibration of the PDH error signal slope at the lock point together with the temperature of the environment furthermore allows a quantification of the vacuum optomechanical coupling rate of the membrane modes (see methods, and \cite{gorodetksy2010determination, saavedra2021tunable}).

\subsection*{Characterization of the optomechanical coupling}
\label{sec:optomechmap}
A linear, dispersive optomechanical coupling manifests as a shift of the optical cavity frequency $\Delta \omega_\text{cav}$ upon a displacement $\Delta z$ of the membrane \cite{thompson2008strong, jayich2008dispersive}. In turn, the cavity field photons act on the pliable membrane through their radiation pressure, closing the interaction loop. The magnitude of this effect strongly depends on the geometry of the MIM system, in particular on the optical intensity on both sides of the membrane. A difference in intensity will cause a non-vanishing net radiation pressure on the membrane resulting in an optomechanical interaction. The interaction strength is given by the vacuum optomechanical coupling rate $g_0 = \frac{\partial \omega_\text{cav}}{\partial z} z_\text{zpf} \coloneqq -G^{(1)} z_\text{zpf}$ with $z_\text{zpf}$ the mechanical resonator's zero-point motion amplitude. As we fix our total cavity length to $\sim \SI{30}{\micro\meter}$, the relevant parameters determining the achievable coupling are the polymer drum support height $h_\text{drum}$ and the membrane thickness $t_\text{drum}$ of the drum (see Fig.~\ref{fig:fig01_generalSetup}~b).
Since the latter is comparable to the probe laser wavelength $\lambda$, it has a strong effect on the local cavity intensity at the membrane surfaces.

\begin{figure}[h!]
\centering
\includegraphics[width=\linewidth]{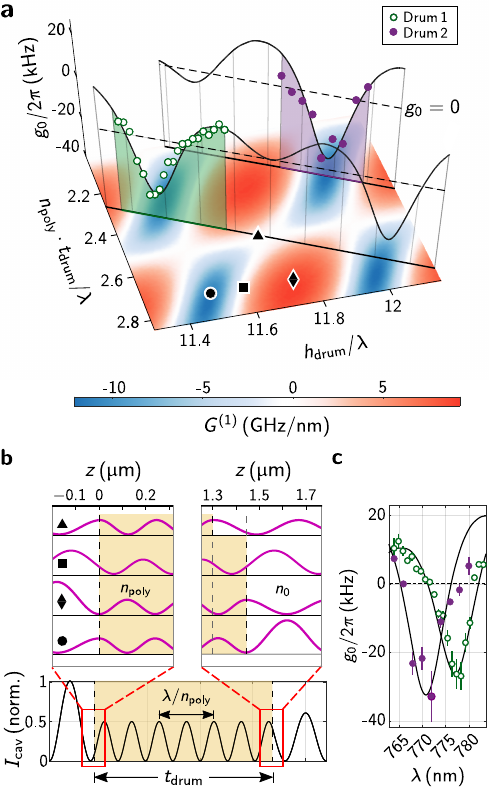}
\caption{
In \textbf{a} the calculated $G^{(1)}(t_\text{drum}, h_\text{drum})$ are shown as a color-map, with \{$t_\text{drum}, h_\text{drum}$\} normalized to the respective material wavelength. Measured $g_0$ for two exemplary polymer drums are highlighted in green (drum $1$) and purple (drum $2$) and plotted along the third plot-axis. By scanning the probe-wavelength, the measurements of the two drums trace out cuts in the map. Four cavity geometries corresponding to special coupling scenarios are highlighted: \mytriangle{black} and $\mysquare{black}$ correspond to $G^{(1)} = 0$,\footref{fn:quadCoup} \mytriangleUSD{black} and \mycircle{black} maximize $|G^{(1)}|$. In \textbf{b} (bottom) a sketch of the cavity intensity $I_\text{cav}$ of an exemplary cavity geometry is shown. The cavity intensity at the two polymer-air/vacuum interfaces is shown for the four scenarios from \textbf{a} (\mytriangle{black}, $\mysquare{black}$, \mytriangleUSD{black}, \mycircle{black}). In \textbf{c} the measured $g_0$ from image \textbf{a} are plotted against the scanned probe laser wavelength $\lambda$ for both drum 1 (green) and drum 2 (purple). }
\label{fig:fig02_lambda}
\end{figure}

We compute the expected frequency-pull factor $G^{(1)}(t_\text{drum}, h_\text{drum})$ in our experiment using two different methods. After finding the resonance condition of the optical cavity field (see methods), the first method considers a small shift $\Delta z$ of the membrane position. Its effect on the cavity resonance frequency as expected from Maxwell's equations is numerically extracted to determine the frequency-pull. For the second method, we employ perturbation theory for Maxwell's equations with moving material boundaries \cite{johnson2002pertubation} (for details, see methods). For this, we use the cavity resonance condition to find the explicit form of the cavity electrical field. The resulting pull-factor (see Eq.~\ref{equation:G1}) scales linearly with the intensity difference on the two membrane-air interfaces and the polymer refractive index $n_\text{poly}$. Both methods yield identical results. The resulting $G^{(1)}(t_\text{drum}, h_\text{drum})$ is shown in Fig.~\ref{fig:fig02_lambda}~a with some highlighted field configurations in b. Maximum frequency-pull factors of up to $\SI{11}{\giga\hertz\per\nano\meter}$ are predicted. The map features a periodic pattern of well-defined minima and maxima with a periodicity of half the (material) wavelength. The asymmetry of positive vs. negative coupling emerges from the two air domains of the cavity not having equal lengths. 

We test our model by measuring $g_0$ for different probe wavelengths on two exemplary polymer geometries: $\{ t_\text{drum 1} \sim \SI{1.2}{\micro \meter},\, h_\text{drum 1} \sim \SI{8.9}{\micro \meter} \}$ and $\{ t_\text{drum 2} \sim \SI{1.15}{\micro \meter},\, h_\text{drum 2} \sim \SI{9.2}{\micro \meter} \}$. The calibrated optical cavity frequency noise spectrum $\mathcal{S}_{\nu \nu}(f)$ is measured (see methods) and the fundamental mechanical resonance is used to extract $g_0$ at ambient conditions from its expected effect on the cavity frequency noise via \cite{gorodetksy2010determination}:

\begin{equation*}
    \mathcal{S}_{\nu \nu}(f)=\frac{2 g_0^2}{4 \pi^2} \cdot \frac{2 \Omega_m}{\hbar} \cdot \frac{2 \Gamma k_B T}{\left(\Omega^2-\Omega_m^2\right)^2+\Gamma^2 \Omega^2}.
\end{equation*}
Here, $f = \Omega/2 \pi$ denotes the noise frequency, $\Omega_m$ the mechanical resonance frequency, $\Gamma/2 \pi$ the mechanical linewidth, $k_B$ Boltzmann's constant and $T$ the temperature of the polymer membrane. The results are shown in Fig.~\ref{fig:fig02_lambda}~a and c. As the probe laser wavelength is scanned, the coupling rate changes through the shift of the cavity intensity on the polymer drum surfaces. This corresponds to the diagonal cuts in Fig.~\ref{fig:fig02_lambda}~a across the coupling landscape for the two representative drum geometries (green -- drum $1$, purple -- drum $2$). Here, tuning the probe laser wavelength from $\SI{760}{\nano\meter}$ to $\SI{785}{\nano\meter}$ is equivalent to a few $\SI{100}{\nano\meter}$ variation of the geometry\footnote{$h_\text{drum}$ is chosen to be $\gtrsim \SI{8}{\micro\meter}$ for these geometries, such that tuning over the complete probe laser wavelength-range changes the cavity field node/anti-node positions by at least half of the (material) wavelength.}. We find a maximum coupling rate of $|g_0|/2\pi = \SI{33\pm 7}{\kilo\hertz}$ at $\lambda = \SI{772}{\nano\meter}$ (see Fig.~\ref{fig:fig02_lambda}~c for drum 2). The sign of $g_0$ in Fig.~\ref{fig:fig02_lambda} is inferred from the fabricated geometry parameters and the expected asymmetry of positive and negative coupling regions.

Four special cases of the intensity distribution are highlighted in Fig.~\ref{fig:fig02_lambda}~a. Their corresponding cavity intensity distribution at the polymer membrane surfaces is shown in Fig.~\ref{fig:fig02_lambda}~b. Vanishing optomechanical coupling is observed for equal intensities on both sides of the membrane, in the simplest case by $t_\text{drum}$ being a multiple of half the material wavelength. Optimal coupling is achieved, if a cavity field node (anti-node) is located at one side of the polymer membrane in combination with a corresponding anti-node (node) at the other\footnote{\label{fn:quadCoup}Note that the quadratic frequency-pull factor is maximized for equal intensities on either side along the connection line of the linear maxima of same $t_\text{drum}$. This corresponds to $\mysquare{black}$ in Fig.~\ref{fig:fig02_lambda}~b. It also vanishes at $t_\text{drum}$ being equal to a multiple of half the material wavelength (\mytriangle{black}). The maximum expected quadratic coupling is $G^{(2)} \gtrsim \SI{200}{\giga\hertz\per\nano\meter\squared}$.}.

\subsection*{Optomechanical spring effect}
We now consider dynamical effects of the intracavity photon number on the mechanical modes of the polymer membrane. As we are working in the fast-cavity/Doppler-regime, where \firstrevision{the cavity decay rate} $\kappa \gg \Omega_m$, the optomechanical spring effect shifts the fundamental flexural mode frequency $\Omega_m$ as 
\begin{equation} \label{eq:springEffect}
    \Delta \Omega_m(\Delta)={g_0}^2 n_\text{cav} \cdot \frac{2 \Delta}{\Delta^2 + \kappa^2 / 4}
\end{equation}
with cavity detuning $\Delta$, cavity photon number $n_\text{cav}$ and vacuum optomechanical coupling rate $g_0$ \cite{aspelmeyer2014cavity}. Effects on the mechanical linewidth are neglected. 

To measure $\Delta \Omega_m$, we make use of a two-tone measurement scheme. The cavity is locked in a PDH feedback loop to a probe laser beam fixed at $\SI{780}{\nano\meter}$ \firstrevision{with low optical power ($<\SI{50}{\micro\watt}$)} (for details, see methods).\COMM{To measure $\Delta \Omega_m$, we make use of a two-tone measurement scheme that utilizes a probe laser beam fixed at $\SI{780}{\nano\meter}$ to which the cavity is locked in a PDH feedback loop (for details, see methods).} $\Omega_m$ is measured from $\mathcal{S}_{\nu \nu}(f)$ as extracted from the calibrated error signal of the feedback loop. To impose a frequency shift of the mechanical resonator, a tunable pump-laser ($\SI{760}{\nano\meter}-\SI{785}{\nano\meter}$) is additionally coupled into the input-port fiber addressing another optical cavity resonance separated by one free spectral range ($\sim \SI{5}{\tera\hertz}$) from the probe. The probe path is decoupled from the pump-laser radiation with a narrow pass-band interference filter. The pump-laser is scanned over the optical resonance modulating the detuning and the intracavity photon number $n_\text{cav}$. During this process, $\mathcal{S}_{\nu \nu}(f)$ is recorded  to extract $\Delta \Omega_m$. \firstrevision{The influence of the low-power probe beam on the measurements is negligible.} The measurements are repeated for a sweep of different pump input powers $P_\text{in}$. 

\begin{figure}[t]
\centering
\includegraphics[width=\linewidth]{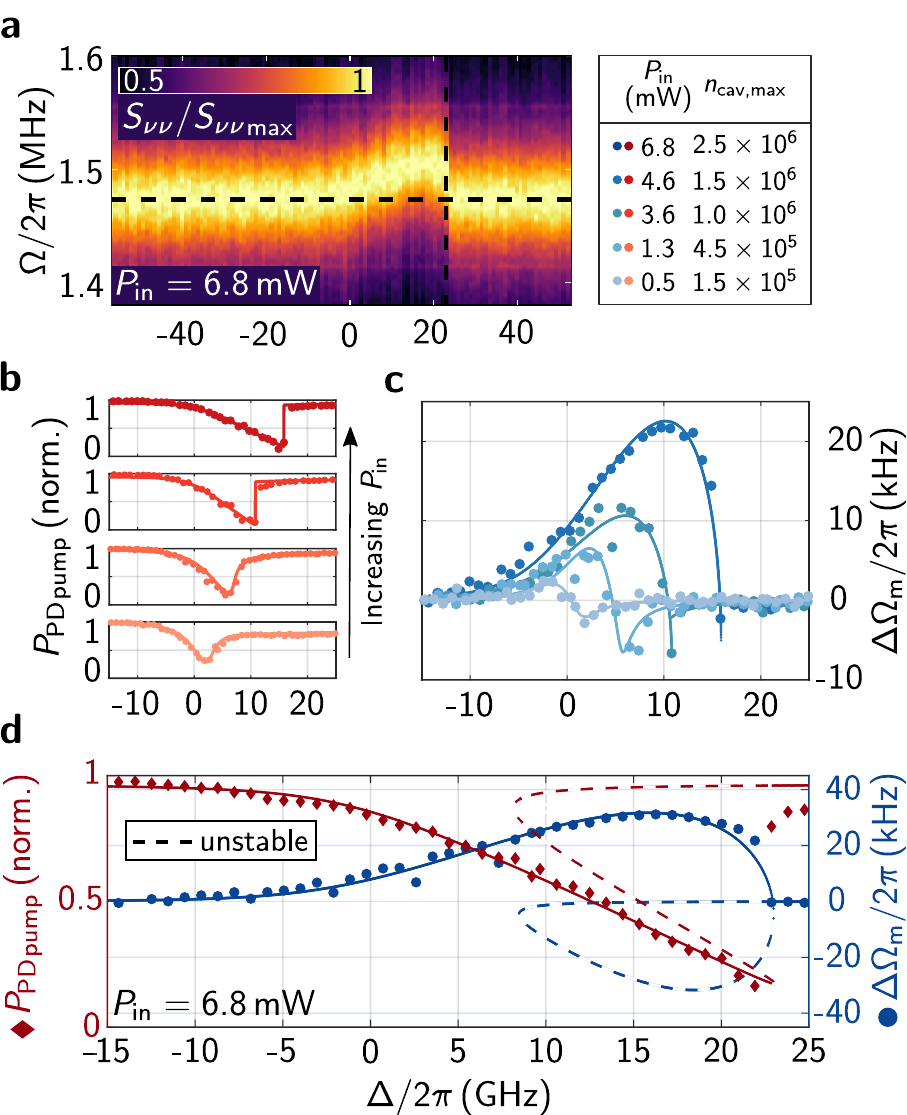}
\caption{\textbf{a} shows the normalized frequency noise spectra $S_{\nu \nu} / S_{\nu \nu \max }$ against cavity detuning $\Delta$ for $P_\text{in} = \SI{6.8}{\milli \watt}$. The dashed black lines highlight the effect of the optomechanical spring on the mechanical resonance frequency $\Omega_m$. The legend in the upper right indicates the corresponding pump powers and \firstrevision{maximum (on-resonance) cavity photon numbers} for \textbf{b}-\textbf{d}. In \textbf{b} the pump-reflection signal ${P_\text{PD}}_\text{pump}$ is plotted against the cavity detuning $\Delta$ for different input powers $P_\text{in}$. In \textbf{c} the corresponding mechanical resonance frequency shift $\Delta \Omega_m$ induced by the optomechanical spring is shown. \textbf{d} shows the normalized pump-reflection signal ${P_\text{PD}}_\text{pump}$ and mechanical resonance frequency shift $\Delta \Omega_m$ induced by the optomechanical spring against cavity detuning $\Delta$ for input power $P_\text{in} = \SI{6.8}{\milli \watt}$. The dashed lines correspond to the unstable or not scanned solutions of Eq.~\ref{equation:ncav}.}
\label{fig:fig03_SpringEffect}
\end{figure}

Fig.~\ref{fig:fig03_SpringEffect}~a shows a typical measurement of $\mathcal{S}_{\nu \nu}(f)$ of the mechanical polymer drum resonance against the cavity detuning $\Delta$ for a comparably large pump input power of $P_\text{in} = \SI{6.8}{\milli\watt}$. At such a power level an additional photothermal nonlinearity of the optical cavity resonance shifts the resonance position during the detuning scan \cite{carmon2004thermal} leading to a discontinuous behavior (black dashed cross) of the signal. The photothermal nonlinearity of the optical mode is also observed in the reflection signal of the pump-laser shown in Fig.~\ref{fig:fig03_SpringEffect}~b. An optomechanical bistability as the origin of this behavior can be excluded as our coupling rate $g_0$ and $n_\text{cav}$ are too small for this effect \cite{aspelmeyer2014cavity}. The nonlinearity is likely related to absorption on the polymer surfaces as stronger nonlinearites are observed in geometries with intensity maxima at the polymer-air interface. 

As the photothermal absorption is slow compared to the optomechanically induced dynamics, we include the shift and bistability of the optical resonance in our analysis by treating it as an additional static detuning of the cavity. The cavity detuning $\Delta$ is therefore modified to $\Delta' = \Delta-\alpha n_\text{cav}$ with $\alpha$ being the photothermal frequency-pull factor. \firstrevision{To find $n_\text{cav}$ for a given detuning, we insert $\Delta'$ in the steady-state solution of the equation-of-motion of the complex light amplitude $\hat{a}$ as obtained from input-output formalism \cite{gardiner1985input}
\begin{equation}
   - \sqrt{\kappa_\text{ex}} \hat{a}_\text{in}= \left(i\Delta'-\kappa/2\right)\hat{a},
 \label{equation:eoma}
\end{equation}
with cavity input coupling $\kappa_\text{ex}$ and complex input field amplitude $\hat{a}_\text{in}$. Squaring and averaging Eq.~\ref{equation:eoma} on both sides leads to a third-order polynomial equation in the cavity photon number $n_\text{cav}$:}
\begin{equation}
   0=\alpha^2 n_{\text {cav }}^3-2 \Delta \cdot \alpha n_{\operatorname{cav}}{ }^2+\left(\Delta^2+\kappa^2 / 4\right) n_{\text {cav }}-\kappa_\text{ex} n_{\text {in }}
 \label{equation:ncav}
\end{equation}
\firstrevision{with cavity photon number $n_\text{cav}=\langle\hat{a}^{\dagger}\hat{a}\rangle$ and input photon number rate $n_\text{in}=\langle\hat{a}_\text{in}^{\dagger}\hat{a}_\text{in}\rangle$}.
The two stable solutions of $n_\text{cav}$ correspond to the photon number expected from the two directions of the detuning scan, whilst the third, unstable solution is not reached \cite{carmon2004thermal}. As shown in Fig.~\ref{fig:fig03_SpringEffect}~b \& d (red curve), we start the scan from $\Delta<0$. As more and more photons are coupled into the cavity, the resonance shifts away from the approaching pump-laser following one of the stable branches for the cavity reflection signal. When the pump-scan catches up with the drifting resonance at the maximum amount of intracavity photons, any further detuning abruptly reduces the thermal frequency-drift of the resonance resulting in a sudden jump to the reflection signal baseline. Using the solutions of Eq.~\ref{equation:ncav}, the extracted cavity photon number $n_\text{cav}(\Delta)$ can be inserted in Eq.~\ref{eq:springEffect} with $\Delta$ being replaced by $\Delta'$. This model is then used to fit the measured data displayed in Fig.~ \ref{fig:fig03_SpringEffect}~c \& d (blue curve). The unstable solution and the not-scanned branch of the reflection signal are included as dashed lines tracing a loop shape. Towards low pump-powers, the transition to a normal dispersive lineshape of the optomechanical spring effect can be observed.
The maximum optomechanical frequency shift measured here was $\Delta\Omega_m/2\pi = \SI{31}{\kilo \hertz}$  for the maximum pump-power that we reached of $P_\text{in} = \SI{6.8}{\milli\watt}$. The cavity finesse in this case was $\mathcal{F}=1400 \pm 300$ and the photothermal frequency-pull factor $\alpha/2\pi=\SI{8\pm 2}{\kilo\hertz}$. Improved cavity finesse by stronger surface polishing, and higher pump-powers will allow us to increase this shift. Currently it already surpasses the mechanical linewidth at cryogenic temperatures. As the shift is on the order of the current frequency disorder of printed polymer drums it could be used to dynamically tune single drums in a multi-mode system into and out of collective resonances.

\subsection*{Mechanical resonance linewidth}
Under ambient atmospheric conditions, the mechanical \firstrevision{quality factor} of the polymer membrane oscillators is limited to $\sim 20$ by about equal parts through damping of the membrane motion by surrounding gas and internal losses of the polymer. In comparison, radiation of mechanical energy into the substrate is negligible due to the impedance mismatch of sound in the polymer and the glass material below. Internal losses of the polymer are characterized by the loss tangent -- the tangent of the phase between imaginary and real part of the dynamic modulus -- of the material and vary strongly between different DLW resists \cite{IPresins}. In addition, the mechanical properties of polymers can exhibit a rich temperature dependence \cite{hartwig1995polymer} with possible secondary glass transitions due to conformation changes of polymer chains. 

\begin{figure}[h]
\centering
\includegraphics[width=\linewidth]{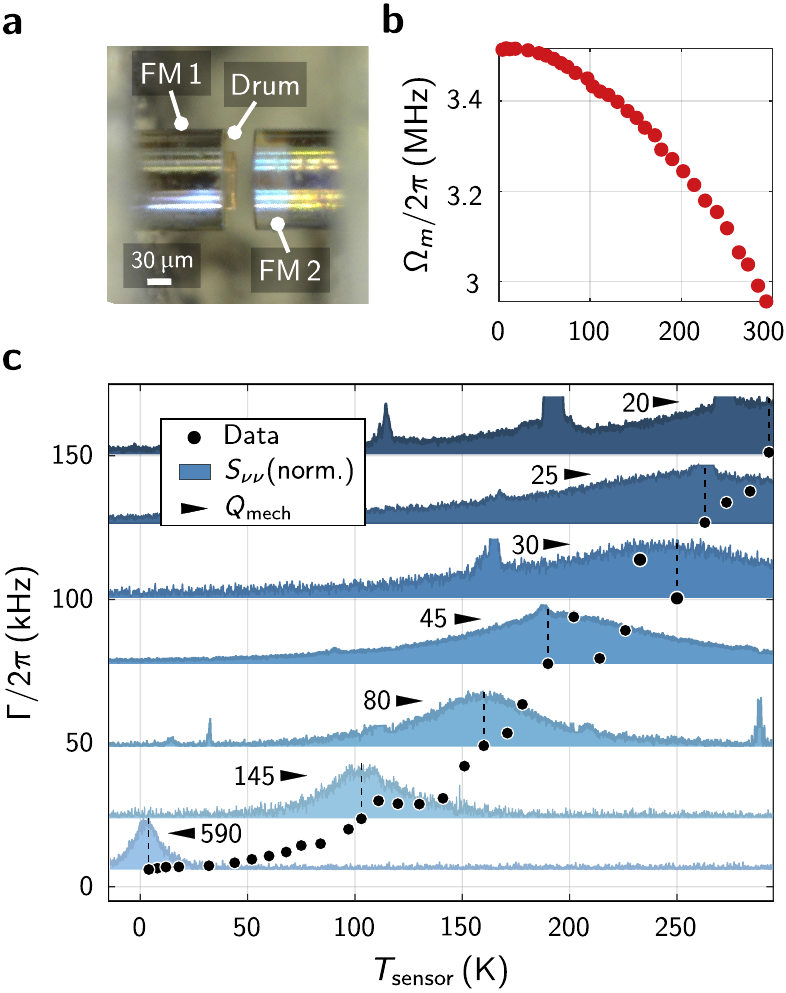}
\caption{Temperature dependence of the mechanical resonator properties. The optomechanical cavity geometry for the experimental setup inside a liquid Helium flow cryostat uses an FFPC with two fiber mirrors (FM$\,1$ and FM$\,2$) inside a slotted glass ferrule that is tunable by an attached piezo \cite{saavedra2021tunable}. The polymer drum is directly fabricated on one of the fiber mirrors as shown in \textbf{a}. Upon cool-down the polymer stiffens causing the resonance frequency to increase as shown in \textbf{b}. In \textbf{c}, the mechanical linewidth at different sensor temperatures is shown. For some exemplary measurements the quality factors and measured spectra are included.}
\label{fig:fig04_LHe}
\end{figure}

To investigate the temperature dependence of the mechanical properties, we place a ferrule-based, passively stable FFPC configuration (for details, see \cite{saavedra2021tunable}) in a liquid Helium continuous flow cryostat. The polymer membrane structure is fabricated on a highly reflective fiber mirror (see Fig.~\ref{fig:fig04_LHe}~a). The sample holder, on which the FFPC is mounted, sits in an evacuated chamber and can be cooled from room temperature down to $\sim\SI{4}{\kelvin}$. The sensor temperature is, however, only a lower bound to the local temperature of the membrane. When cooling down the membrane resonator, the polymer material stiffness and pre-strain increase, which leads to an increase of the mechanical resonance frequency from about $\SI{3}{\mega\hertz}$ to $\sim\SI{3.5}{\mega\hertz}$ (see Fig.~\ref{fig:fig04_LHe}~b). At the same time the mechanical linewidth drops from about $\SI{150}{\kilo\hertz}$ down to $\sim\SI{6}{\kilo\hertz}$ corresponding to a mechanical quality factor of $\sim 600$ as shown in Fig.~\ref{fig:fig04_LHe}~c. Kinks in the temperature dependence of the mechanical linewidth may indicate possible secondary glass transitions~\cite{hartwig1995polymer}. \COMM{Scattering of high energy phonons is another loss mechanism that is suppressed at low temperatures.}

The remaining loss, decoherence or linewidth broading of the polymer membrane resonators at cryogenic temperature can be caused by various mechanisms. A possible candidate is a mechanical coupling to \COMM{a bath of effective two-level systems corresponding to }the conformation changes of polymer molecule chains. Also scattering of high energy phonons can appear, which would be suppressed at even lower temperatures. Due to the comparably large thickness of the membrane resonators, thermo-elastic dissipation \cite{zener1938internal, serra2012inhomogeneous} is expected to play another major role. 
Engineering the mechanical mode to reduce strain gradients and further polishing using oxygen plasma ashing to thin down the membrane will help to reduce \thirdrevision{these loss channels} and can be combined with isolation and soft-clamping techniques to further increase the mechanical quality factor \cite{tsaturyan2017ultracoherent, fedorov2019generalized}. 

\secondrevision{Pre-straining\COMM{, as induced by the usual shrinkage of the polymeric resist during polymerization (see methods)}, together with a thinned membrane, will further help to dilute dissipation and improve the mechanical quality factor\COMM{ by several orders of magnitude}.}\COMM{\firstrevision{Anticipating similar improvements of $Q_\text{mech}/Q_\text{int} \sim 10^4$ by dissipation dilution, as regularly achieved on other platforms \cite{tsaturyan2017ultracoherent}, would allow for mechanical quality factors exceeding one million using the same DLW resist material.}} \thirdrevision{An advantageous resource for pre-straining the material will be the usual shrinkage of the polymeric resist during polymerization. The comparably large tolerance of polymers to straining that is on par or even exceeding the limit strain of conventional materials used for micro-mechanical oscillators can enable comparably large dilution factors partly compensating the lower intrinsic mechanical quality factor. According to the manufacturer, the employed IP-S resist can show shrinkage between 2-\SI{12}{\percent} under tuned fabrication conditions that were not explored here (other sintered glass-composite resists even up to \SI{26.7}{\percent}). Due to the anchoring of the structure the shrinkage in the development directly translates into strain. Assuming \SI{10}{\percent} strain, a dilution factor of $\sim 280$ would be reached for a \SI{100}{\micro\meter} long, $\lambda/4$--thick (\SI{195}{\nano\meter}) structure \cite{fedorov2019generalized, villanueva2014evidence, gonzalez1994brownian, manjeshwar2023high}. For the cryogenic intrinsic quality factor this would result in a $Q_\text{mech} >\SI{1.5e5}{}$ even without further engineering of the resonators. Other resists with higher intrinsic quality and even stronger shrinkage can potentially boost this prospect by another order of magnitude.}

The exploration of the material properties of other 3D DLW resists under cryogenic conditions will \firstrevision{also} help to further identify low-loss materials. Moreover, 3D DLW structures can be used as blanks for subsequent material deposition, where the resist blank is later removed by oxygen plasma ashing or suit as stamp frames (e.g. from PDMS) in combination with other membranes or 2D materials like transition metal dichalcogenides. 

\section{Discussion}
We have demonstrated an FFPC-integrated optomechanical membrane-in-the-middle experiment with 3D direct laser-written membrane structures. \firstrevision{Despite the relatively low membrane reflectivity, large optomechanical couplings of $g_0\sim\SI{30}{\kilo\hertz}$ are realized since the comparatively large membrane thickness \cite{thompson2008strong, flowers2012fiber, shkarin2014optically, fogliano2021mapping, rochau2021dynamical} of several $\lambda/4$ allows us to maximize the intensity differences on the two membrane-air interfaces. Reducing the effective membrane mass by thinning the membrane thickness down to the order of a single $\lambda/4$ thickness will further enhance the optomechanical coupling.}

The unconventional polymer material stemming from the 3D DLW fabrication process results in moderate intrinsic mechanical quality factors at ambient conditions. However, a significant gain in the mechanical Q-factor is realized at cryogenic temperatures. Furthermore, the large range of specialized 3D DLW materials \cite{IPresins}, the possible implementation of dissipation dilution techniques \cite{fedorov2019generalized}, the use of 3D DLW structures as blanks for material growth, and their capabilities as stamping frames will allow a fast advancement of the presented platform. In turn, optomechanical low-temperature experiments enable highly sensitive material characterization of DLW materials, which will contribute to the current rapid development of specialized resists. 

\begin{figure}[h]
\centering
\includegraphics[width=\linewidth]{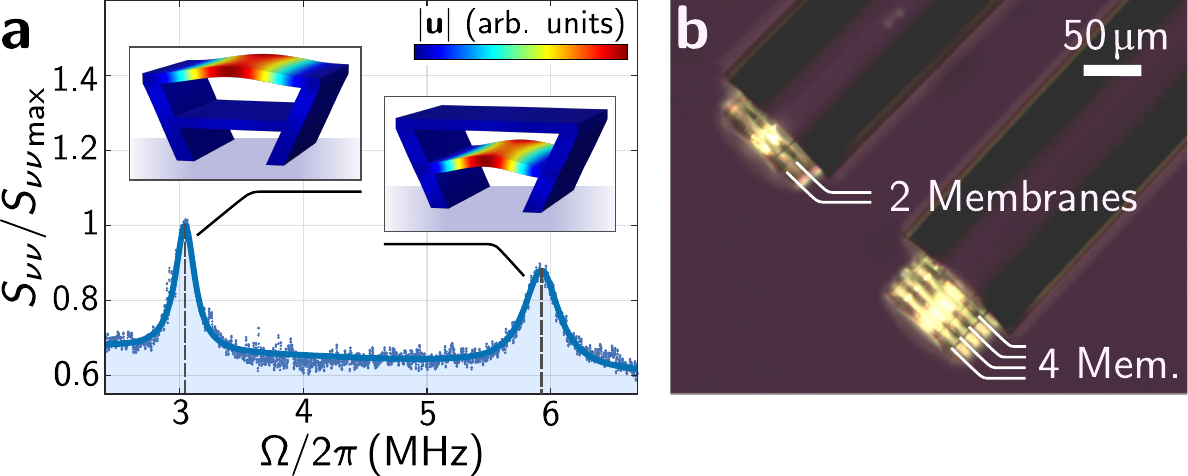}
\caption{\thirdrevision{\textbf{a} shows the optical noise spectrum of a cavity with two mechanical membrane modes and their corresponding simulated displacement fields. The modes in this example structure are deliberately detuned by tilting the supports and thereby the size of the corresponding membrane. \textbf{b} Showcase example of multi-mode mechanical structures realized using direct laser-writing showing stacks of multiple, freely suspended drums (here 2 and 4) that are directly fabricated on the tip of an optical fiber.}}
\label{fig:membraneStacks}
\end{figure}

The large flexibility of the 3D DLW fabrication allows the combination of this highly integrated platform with additional structures like electrodes \cite{pfeifer2022achievements} for electromechanical coupling, or emitters, and makes it scalable towards multi-membrane experiments. These can be both realized as membrane-stacks in a single FFPC \thirdrevision{that are considered to lead to improved values of the optomechanical coupling} \cite{bhattacharya2008multiple, xuereb2012strong}\COMM{ , e.g. on a fiber mirror\firstrevision{(see methods)}}, or as a planar 2D mechanical metamaterial on a DBR substrate. \thirdrevision{First experiments, where we introduce stacks of two mechanical membranes in a single optical cavity, see Fig.~\ref{fig:membraneStacks}, show the feasibility of this approach. In contrast to other platforms \cite{gaertner2018integrated, manjeshwar2023free} the number of layers in such stacks can easily be extended beyond two and the geometry of each membrane separately adjusted to bring multiple membranes in tune.} \firstrevision{Using DLW, mechanical device layers can also be added and combined with a plethora of other platforms for light-matter interaction. Acoustic metamaterials in such device layers} would benefit from the large optomechanical spring effects allowing for tunable mechanics and optical reconfiguration of mechanical multi-mode circuits with much less elaborate tuning techniques than required in other optomechanical platforms \cite{pfeifer2016design}. This will enable vibration-routing in 2D metamaterials, distributed sensing in multi-mode mechanical structures, and fiber-tip-integrated sensing of motion and force. 

\section*{Methods} \label{sec:meth}
\subsection*{Sample fabrication} \label{sec:sampleFab}
The polymer drum structures were fabricated using a commercial 3D lithography system (Nanoscribe Photonic Professional GT+, Nanoscribe GmbH \& Co. KG, Germany) that uses two-photon-polymerization. A $63\times$ objective was used in combination with the photoresist IP-S in dip-in configuration to print on the end-facet of the single-mode fiber or DBR substrates using the system's piezo-mode. The system features a femtosecond laser centered at 780 nm that patterns and polymerizes the resist at the objective's focus. The hatching distance for the feet and frame were chosen comparably coarse, while the membrane was patterned fine ($\SI{65}{\nano\meter}$) using a laserpower of $22\,\%$ (corresponding to $\SI{11}{\milli\watt}$) and a scanspeed of $\SI{50}{\micro\meter/s}$. The unpolymerized resist part was removed via immersion in propylene glycol methyl ether acetate (PGMEA) ($\SI{30}{min}$) and a subsequent bath in isopropyl alcohol ($\SI{30}{min}$). In a post-development step, the structure was flood exposed with a $\SI{6}{\watt}$ UV-lamp while sitting in isopropyl alcohol ($\SI{10}{min}$). \thirdrevision{Whilst this procedure can minimize structure deformation during the development, it also only results in minimal shrinkage of the resist. Under certain fabrication conditions IP-S is specified to exhibit 2-\SI{12}{\percent} shrinkage. For comparison, the silica-compound based GP-Silica resist can even show up to \SI{26.7}{\percent} shrinkage in the sintering step alone.} Further post-processing of the membrane was performed using oxygen plasma-ashing for surface polishing to reduce scattering losses from the interfaces. The polymer structures were fabricated on the DBR substrate or fiber mirror that constitute the higher reflective cavity mirror ($\SI{10}{ppm}$ transmission). \firstrevision{To reduce the complexity of the analysis and to focus on the fundamental material properties of the resist, the comparably simple rectangular membrane design is presented in this article. However, using this process, several different geometries were realized and much more complex structures including for example membrane stacks as shown in Fig.~\ref{fig:membraneStacks} can be fabricated.} \thirdrevision{The thinnest membrane size in such structures can be reduced below the voxel height $h_\text{v}$ (smallest voxel: $ (\diameter_\text{v},h_\text{v}) \sim (0.2,0.7)\, \SI{}{\micro\meter}$) by oxygen plasma-ashing. The accuracy of the placement of the membranes is given by the piezo stage positioning accuracy (\SI{10}{\nano\meter} in the full $\SI{300}{\micro\meter}\times\SI{300}{\micro\meter}$ writing field),  which is considerably smaller than $\lambda/4$ as required for the membrane positioning. Possible tilts $\alpha_\text{tilt}$ of the structure are corrected by the automatic interface finder, but would also only play a role for tilts approaching $\alpha_\text{tilt}\sim (\lambda/4)/L_\text{mem}$ with $L_\text{mem}$ the length of the membranes. Detailed characterizations of such multi-membrane devices will be presented in future work.}\COMM{\secondrevision{Shrinkage of photoresists during polymerization or sintering (e.g. IP-S: 2-\SI{12}{\percent}, or GP-Silica: \SI{26.7}{\percent} during sintering) can furthermore be exploited to implement dissipation dilution techniques for pre-strained materials \cite{fedorov2019generalized, villanueva2014evidence} that for realistically achievable geometries can boost the mechanical quality factor by several orders of magnitude.}\firstrevision{Detailed characterizations of such devices will be presented in future work.}}

\subsection*{Experimental setup} \label{sec:expSetup}

\begin{figure*}[ht]
\centering
\includegraphics[scale=1.0]{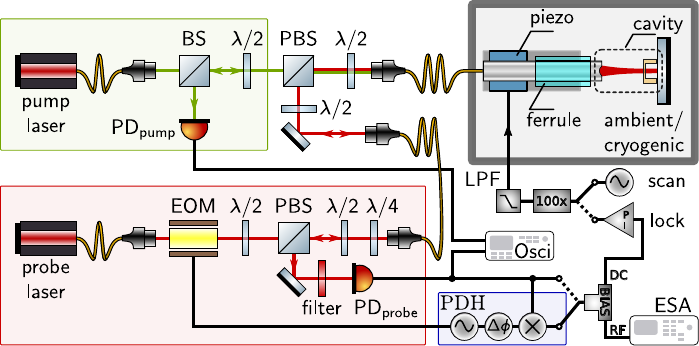}
\caption{Overview of the experiment setup. The green highlighted pump laser is only used within the optomechanical spring measurements. The optical cavity with the mechanical membrane resonator (gray box) is either located in ambient conditions (single fiber mirror and DBR substrate) or situated on the sample holder of a liquid Helium flow cryostat (two fiber mirrors in a glass ferrule). The optical cavity length can either be scanned or locked to the probe laser (highlighted in red) using either a Pound-Drever-Hall-type (highlighted in blue) or a side-of-fringe lock. For this purpose and as a frequency-meter in case of a scanned cavity, an electro-optic modulator is used to create sidebands to the main probe laser tone at adjustable RF-frequency up to $\SI{12}{\giga\hertz}$. The photodiode reflection signals and the electrical noise spectrum of the lock can be measured using conventional oscilloscopes and an electric spectrum analyzer.}
\label{fig:experimentSetup}    
\end{figure*}

The full experiment setup is sketched in Fig.~\ref{fig:experimentSetup}. The optical mode is characterized by measuring the reflected optical power of a fixed wavelength probe laser (probe setup red highlighted in Fig.~\ref{fig:experimentSetup}, see also \cite{gallego2016high, saavedra2021tunable}) on a photodiode (PD$_\text{probe}$), while the cavity length is scanned using a piezo-electric element that is glued to the optical fiber. The high-voltage drive of the piezo is generated by a $100 \times$ voltage amplifier and a subsequent low pass filter (LPF) to reject high-frequency electric noise. The alignment of the fiber during the scan process is secured by a glass ferrule ($\SI{131}{\micro\meter}$ bore diameter), in which the optical fiber can slide. The incoming and reflection signal are split on a polarizing beam splitter (PBS). The subsequent waveplates ($\lambda/2$ and $\lambda/4$) are used to adjust the polarization of the reflected light to be perpendicular to the polarization of the incoming laser beam causing the PBS to direct most of the reflection onto PD$_\text{probe}$. An electro-optic modulator (EOM) adds sidebands to the main probe laser tone. The reflection signal measured with the oscilloscope therefore features three reflection dips with a spacing of the RF-drive of the EOM that is thereby used as a frequency-meter. 

To characterize the optomechanical coupling strength the optical cavity is locked to the probe laser tone using a Pound-Drever-Hall (PDH, blue highlighted) or side-of-fringe (SoF) lock. The slope of the lock signal is calibrated using the frequency-meter enabled by the EOM-sidebands. Using this, the measured voltage noise of PD$_\text{probe}$ can be converted to the optical cavity frequency noise. As the bandwidth of the feedback loop is small ($\sim \SI{1}{\kilo\hertz}$) compared to the mechanical resonance frequency ($\gtrsim \SI{1}{\mega\hertz}$) a bias tee is used to split the low from the high-frequency component. The DC-like part is used in the feedback loop, whilst the high-frequency part is analyzed using an electric spectrum analyzer (ESA). Mechanical mode frequency and linewidth can directly be extracted from there. The vacuum optomechanical coupling strength is then inferred by comparing the optical cavity frequency noise with the expected thermal noise of a mechanical oscillator of the measured frequency and linewidth at ambient temperature conditions \cite{gorodetksy2010determination}.

In order to measure the dynamic optomechanical spring effect a second pump laser is used (highlighted in green), while the optical cavity remains tightly locked to the probe laser. The pump laser addresses a second optical cavity mode separated by one free spectral range towards higher frequencies from the probe-cavity-resonance. For variable pump laser powers, the pump frequency is scanned. The reflection signal of the pump is recorded on a second photodiode (PD$_\text{pump}$), while the locked probe laser error signal is used for the characterization of the mechanical mode frequency. 

\firstrevision{ 
\subsection*{Optical properties of the membrane-cavity system}}
The FFPC designs used in this work are based on either a hemi-cavity design with a fiber mirror that can scan structures on a flat, highly reflective substrate inspired by \cite{mader2015scanning}, or a passively stable, monolithic FFPC realization with two fiber mirrors inside a glass ferrule as demonstrated in \cite{saavedra2021tunable}.

\firstrevision{ 
We use a standard $\ce{CO2}$ laser ablation system to machine spherical-like depressions onto the center of single-mode optical fibers \cite{hunger2010fiber}. The resulting concave indentation on our fiber end-facets features typical radii of curvature of $\sim\SI{200}{\micro \meter}$ with usable spherical diameters of $>\SI{40}{\micro\meter}$. The prepared fiber-tips are coated with alternating layers of $\ce{Ta2O5}$ and $\ce{SiO2}$\COMM{ with a layer thickness of $\lambda/4$}, resulting in highly reflective fiber end-facets for wavelengths ranging from $750 -\SI{800}{\nano\meter}$ at $\text{AOI}=\SI{0}{\degree}$. These fiber mirrors make up the cavity geometry and are coated for high transmission ($\SI{2000}{ppm}$) to be used as in-coupling fiber mirrors to the cavity and low transmission ($\SI{10}{ppm}$) blank mirrors for interfacing the direct laser-written membranes. The empty cavity finesse of $\sim 2800$ (no polymer membrane inside cavity volume) is primarily determined by the transmission losses of the cavity mirrors.}

\begin{figure}[h]
\centering
\includegraphics[width=0.9\linewidth]{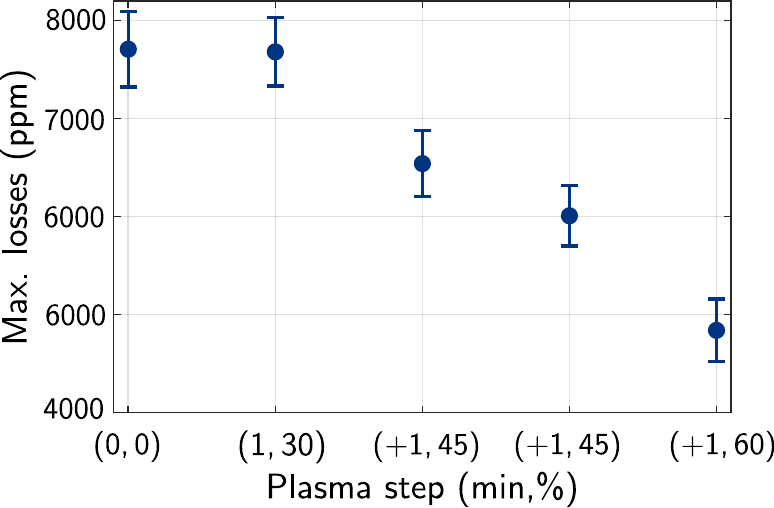}
\caption{\firstrevision{Effect of oxygen plasma polishing on the amount of surface scattering from a DLW-fabricated thin film. The cavity losses are determined by measuring the minimal optical finesse of a polymer block structure with a height variation placed directly on a mirror substrate. Evaluating at the thickness of minimal finesse thereby ensures an intensity maximum of the intra-cavity field at the scattering polymer interface. The reduction of losses can therefore be attributed to the surface losses only, whilst their absolute value contains additional losses from the imperfections in the large bulk block below. In each step the same polymer structure underwent subsequent plasma steps of duration $t_\text{polish}$ (minutes) at plasma power $P_\text{plasma}$ (\%) yielding the horizontal axis sets $(t_\text{polish},P_\text{plasma})$.}}
\label{fig:O2plasmaPolish}
\end{figure}

\firstrevision{ 
The optical losses induced by the polymer membrane at lower input laser powers ($<\SI{500}{\micro \watt}$) are dominated by surface scattering effects at the polymer-air interfaces. Depending on the cavity field distribution, optical scattering losses induced by the polymer membrane can range from $\sim \SI{5000}{ppm}$ with cavity intensity maxima on both polymer-air interfaces to almost fully recovering the empty cavity optical quality in the case of intensity minima on both interfaces. The scattering from the polymer surfaces can be strongly reduced by polishing through oxygen plasma\footnote{Plasma cleaner system: Zepto-BR-200-PCCE}, which smoothens the surface as shown in Fig.~\ref{fig:O2plasmaPolish}. As the plasma removes material from the surface, this technique can also be used to reduce the thickness of the membrane. Another possibility to in the future enhance the optical properties is to combine the fabrication with an atomic layer deposition based uniform growth that can flatten the surface even further.}

\firstrevision{ 
At higher input laser powers ($>\SI{500}{\micro \watt}$), additional photothermal nonlinearities of the optical cavity resonance become relevant. Since we observe stronger nonlinearities in cavity geometries with intensity maxima at the polymer-air interfaces, these effects can most likely be attributed to surface absorption as for example induced by not-passivated bonds of the polymer. The additional absorption will then lead to heating and mechanical expansion of the membrane, causing the additional static photothermal detuning as observed in the experiments.}

\subsection*{Coupling calculations}
To compute the expected optomechanical frequency-pull factor $G^{(1)}(t_\text{drum}, h_\text{drum})$ for a particular drum thickness and support height, we use two methods as described in section \ref{sec:optomechmap}. For both methods, we first numerically find the set of $\{l_\text{drum},t_\text{drum},h_\text{drum}\}$ that fulfils the resonance condition of the optical cavity with the probe light at \thirdrevision{$\lambda={2 \pi}/{k_0}$}. Here, $h_\text{drum}$ is the polymer drum support height, $t_\text{drum}$ the polymer membrane thickness, and $l_\text{drum} = L_\text{cav} - t_\text{drum} - h_\text{drum}$, where, $L_\text{cav}$ denotes the separation between the two cavity mirrors (for an overview see Fig.~\ref{fig:fig01_generalSetup}~b). For each pair $\{t_\text{drum},h_\text{drum}\}$, a $l_\text{drum}$ at $L_\text{cav} \sim \SI{30}{\micro\meter}$ can be chosen to match the resonance condition $R(l_\text{drum},t_\text{drum}, h_\text{drum}) = 0$, with (suppressing subscripts):
\begin{equation}
\label{equation:rescond}
R\left(l, h, t\right)=A_{+}\left(l\right) \cdot e^{-i n_\text{poly} k_0 t}-A_{-}\left(h\right) \cdot e^{i n_\text{poly} k_0 t}
\end{equation}
and with coefficients $A_{\pm}(z)$:
\begin{equation*}
    A_{\pm}(z)=\frac{1 \mp i n_\text{poly} \cdot \tan \left(k_0 z\right)}{1 \pm i n_\text{poly} \cdot \tan \left(k_0 z\right)} \; .
\end{equation*}

In the picture used here, the cavity consists of a loss-less dielectric membrane with refractive index $n_\text{poly}$. It is placed between two mirrors with perfectly conducting surfaces, enforcing field nodes at their positions (the penetration depth of the DBR will in reality lead to a slightly modified resonance cavity length). As the Rayleigh length $z_R \gg L_\text{cav}$, the Gaussian-beam properties of the cavity field are neglected and a simple standing-wave ansatz is made. There, the tangential component of the electric field at both mirror surfaces needs to vanish and the tangential components of both the electric- and magnetic fields need to be continuous at the dielectric interfaces \cite{ujihara2010resonance}. Applying these conditions to the cavity field, we numerically find the condition that fixes the $\{l_\text{drum},t_\text{drum},h_\text{drum}\}$ - triplet. This is now used to extract the optomechanical frequency-pull factor $G^{(1)}(t_\text{drum}, h_\text{drum})$: 

In \textbf{Method 1}, we use Eq.~\ref{equation:rescond} by applying a small shift to the membrane, altering the cavity geometry to $l_\text{drum} \rightarrow l_\text{drum} + \Delta z$, $h_\text{drum} \rightarrow h_\text{drum} - \Delta z$ and $k_0 \rightarrow k_0 + \Delta k$. This simulates the effect of the linear displacement of the membrane center region on the cavity resonance condition\footnote{For this mathematical treatment, $h_\text{drum}$ does not directly correspond to the polymer drum support height, but more accurately to the distance between the displaced membrane at its center and the subjacent mirror. The mode width of the optical mode is sufficiently smaller than the effective membrane radius to ensure that the optical mode only overlaps with an almost constant displacement of the drum membrane .}. Inserting the shifted geometry into Eq.~\ref{equation:rescond} allows to numerically extract a $\Delta k$ that again fulfils Eq.~\ref{equation:rescond} and thereby reflects the shift on the resonance frequency $\Delta \omega_\text{cav}$. The ratio between the frequency shift and the small displacement allows us to numerically determine the frequency-pull $G^{(1)}(t_\text{drum}, h_\text{drum})$. Aside from the linear frequency-pull factor, higher orders can also be numerically evaluated. E.g. for the quadratic pull-factor values of $\gtrsim \SI{200}{\giga\hertz\per\nano\meter\squared}$ are expected.

For \textbf{Method 2}, we use the optical field that fulfills the resonant cavity condition and apply perturbation theory to find the shift of the optical resonance associated with a shift of the dielectric boundaries of our geometry as derived by Johnson et al. in \cite{johnson2002pertubation}. Here, Maxwell’s equations are written as an eigenproblem of the electric cavity field\footnote{For convenience, the basis-independent representation of the electric field as "Bra"- and "Ket"-vectors is utilized, with inner product ${\left\langle E \mid E^{\prime}\right\rangle \equiv \int \mathbf{E}^* \cdot \mathbf{E}^{\prime} \mathop{dV}}$.} $|E\rangle$ with eigenfrequency $\omega_\text{cav}$ given by the well-known source-free wave equation with overall cavity permittivity $\epsilon(z)$:
\begin{equation}
\label{equation:eigen}
\nabla^2|E\rangle=\left(\frac{\omega_{\text {cav }}}{c}\right)^2 \epsilon(z)|E\rangle.
\end{equation}

Due to the vibrational motion of the
drum membrane, the effective dielectric permittivity of the cavity geometry experiences a local shift $\mathop{\delta\epsilon}$ due to a perturbative shift in position of the drum membrane $\mathop{\delta z}$. We expand $|E\rangle$, $\omega_\text{cav}$ to first-order in $\mathop{d z}$. Plugging these expressions back into Eq.~\ref{equation:eigen} and neglecting terms of $\mathcal{O}\left(\mathop{d z}^2\right)$ leads to the first-order correction of the resonator frequency in differential form \cite{johnson2002pertubation}:
\begin{equation}
\label{equation:firstorder}
G^{(1)}=\frac{d \omega_{\mathrm{cav}}^{(1)}}{d z}=-\frac{\omega_{\mathrm{cav}}^{(0)}}{2} \frac{\left\langle E^{(0)}\left|\frac{d \epsilon}{d z}\right| E^{(0)}\right\rangle}{\left\langle E^{(0)}|\epsilon| E^{(0)}\right\rangle}
\end{equation}\\
We then make use of the explicit parametrization of the cavity \thirdrevision{permittivity}
\begin{equation*}
\epsilon(z)=\epsilon_{2}+\Delta\epsilon \big(\Theta(z-z_0) - \Theta(z-(z_0+t_\text{drum}))\big)\; ,
\end{equation*}
with $\Delta \epsilon = \epsilon_1-\epsilon_2$ ($\epsilon_1$: polymer, $\epsilon_2$: air/vacuum) \thirdrevision{and the Heaviside step function $\Theta(z)$}. The first drum surface $S_\text{L}$ is located at $z_0$ and the second drum surface $S_\text{R}$ at $z_0+t_\text{drum}$ following the coordinate conventions of Fig.~\ref{fig:fig02_lambda}~b.

Inserting this expression back into Eq.~\ref{equation:firstorder}, we arrive at the explicit form of the optomechanical frequency-pull $G^{(1)}$:
\begin{equation}
\label{equation:G1}
G^{(1)}=\frac{\mathop{d \omega_{\mathrm{cav}}}}{\mathop{d z}}=\frac{\omega_{\mathrm{cav}}^{(0)}}{2} \frac{\int_{S_R} \Delta \epsilon\left|\mathbf{E}_{\|}^{(0)}\right|^2 \mathop{d S} -\int_{S_L} \Delta \epsilon\left|\mathbf{E}_{\|}^{(0)}\right|^2 \mathop{d S}}{\int_V \epsilon(z)\left|\boldsymbol{E}_{\|}^{(0)}\right|^2 \mathop{d V}} \; ,
\end{equation}
with the total resonator volume $V$ and parallel electric field component $\mathbf{E}_{\|}^{(0)}$ (explicit form given by Eq.~\ref{equation:rescond}). 
As detailed in section \ref{sec:optomechmap}, Eq.~\ref{equation:G1} allows us to read-off the frequency-pull $G^{(1)}(t_\text{drum}, h_\text{drum})$ for a specific cavity geometry by considering the cavity field distribution on both surfaces of the polymer drum. It also provides the physical interpretation of the coupling to be caused by differing field intensities on the membrane-air interfaces. The coupling landscape retrieved from method 1 and method 2 agree for all tested geometries. For our cavity geometry of interest, the results are shown in Fig.~\ref{fig:fig02_lambda}~a. 

\subsection*{Finite element simulations} \label{sec:femSimulations}
We use COMSOL Multiphysics\textregistered ~\cite{comsol} to perform finite element simulations of the mechanical resonator structures. The simulated geometry consists of the polymer membrane (IP-S: Dynamic modulus $E = \complexqty{5.33+0.26 i}{\giga\pascal}$, Poisson's ratio $\nu=0.3$, density $\rho = \SI{1.15}{\kilo\gram\per\cubic\deci\meter}$;~\cite{IPresins}) with a silica substrate below. An outer shell of the substrate is defined as a perfectly matched layer (PML) to implement radiation losses into the substrate (see Fig.~\ref{fig:fig01_generalSetup}~b). Whilst these radiation losses are not the dominating loss mechanism at room temperature, they are strongly dependent on the polymer geometry and can be reduced to limiting quality factors of $\gg 10^6$ for optimized support dimensions and with included isolation cuts. By reducing the internal polymer losses, e.g. through other resist materials, high quality factor mechanical resonators would be feasible. Aside from the quality factor and resonance frequency, we use the simulation to extract the effective mass of the fundamental flexural mode (on the order of $\sim\SI{2.4}{\nano\gram}$ depending on the specific geometry) and the corresponding zero point motion ($\sim\SI{3.2}{\femto\meter}$) \cite{aspelmeyer2014cavityBook}.

\bibliography{references} 
 
\section*{Acknowledgements}
The authors would like to thank Prof. Dieter Meschede for supporting the first experiments on this project. The authors acknowledge funding by the Deutsche Forschungsgemeinschaft (DFG, German Research Foundation) under Germany's Excellence Strategy – Cluster of Excellence Matter and Light for Quantum Computing (ML4Q) EXC 2004/1 – 390534769 as well as funding from the Bundesministerium für Bildung und Forschung (BMBF, Federal Ministry for Education and Research) - project FaResQ. S.H. furthermore acknowledges funding from the European Union's Horizon 2020 program under the ERC consolidator grant RYD-QNLO (Grant No. 771417).\\

\section*{Author contributions statement}
H.P., L.T., A.F., and S.L. came up with the concept and planned the experiments. A.F. and L.T. performed the fabrication and L.T. conducted the optical measurements. L.T., A.F., and H.P. set up the numerical simulations and analyzed the data. All authors contributed to the writing of the manuscript.\\

\section*{Data availability}
The data of this study is available from the corresponding author (H.P.) upon reasonable request.\\

\section*{Competing interests}
The authors declare no competing interests.

\end{document}